\documentclass[preprintnumbers,amsmath,amssymb,floatfix,11pt,prd,onecolumn,
superscriptaddress,nofootinbib]{revtex4}
\usepackage{latexsym}
\usepackage{epsfig}
\usepackage{epstopdf}
\usepackage{amssymb}

\begin{document}

\title{\bf Timelike geodesics of a modified gravity black hole immersed in an axially symmetric magnetic field }

\author{ Saqib Hussain }
\email{s.hussain2907@gmail.com}\affiliation{Department of Physics, School of Natural
Sciences (SNS), National University of Sciences and Technology
(NUST), H-12, Islamabad, Pakistan}

\author{Mubasher Jamil}
\email{mjamil@sns.nust.edu.pk; jamil.camp@gmail.com}\affiliation{Department of Mathematics, School of Natural
Sciences (SNS), National University of Sciences and Technology
(NUST), H-12, Islamabad, Pakistan}

\begin{abstract}
{\bf Abstract:} We  investigate  the dynamics of a neutral and a
charged particle around a black hole in modified gravity immersed in
magnetic field. Our focus is on the scalar-tensor-vector theory as
modified gravity. We are interested to explore the conditions on the
energy of the particle under which it can escape to infinity after
collision with another neutral particle in the vicinity of the black
hole. We calculate escape velocity of particle orbiting in the
innermost stable circular orbit (ISCO) after the collision. We study
the effects of modified gravity on the dynamics of particles.
Further we discuss how the presence of magnetic field in the
vicinity of black hole, effects the motion of the orbiting particle.
We show that the stability of ISCO increases due to presence of
magnetic field. It is observed that a particle can go arbitrary
close to the black hole due to presence of magnetic field.
Furthermore ISCO for black hole is more stable as compared with
Schwarzschild black hole. We also discuss the Lyapunov exponent and
the effective force acting on the particle in the presence of
magnetic field.

\end{abstract}
 \maketitle
\newpage
\section{Introduction}

Theories of modified gravity (such as $f(R)$ theory, Lovelock
gravity, Gauss-Bonnet theory etc) are constructed by adding
curvature correction terms in the usual Einstein-Hilbert action
through which the cosmic accelerated expansion might be explained
\cite{Dark} (see also \cite{rev} for reviews on modified gravity).
Such correction terms give rise to solutions of the field equations
without invoking the concept of dark energy. To find the dynamical
equations one can vary the action according to the metric. There is
no restriction on the gravitational Lagrangian to be a linear
function of Ricci scalar $R$ \cite{55}. Recently some authors have
taken into serious consideration the Lagrangians that
are``stochastic" functions with the requirement that it should be
local gauge invariant \cite{56}. This mechanism was adopted in order
to treat the quantization on curved spacetime. The result was that
corrective term in the Einstein Hilbert Lagrangian arises due to
either background geometry and interactions among quantum fields or
gravitational self interaction \cite{57}. Furthermore, it is also
realized that such corrective terms should be incorporated if one
wants to obtain the effective action of quantum gravity on Planck
scale \cite{58}. Besides fundamental physical motivation, these
theories have acquired a huge interest in cosmology as they exhibit
inflationary behaviors and that the corresponding  cosmological
model seem very realistic \cite{60,61}. In this article, our focus
will be on the scalar-tensor-vector theory (will be referred as MOG)
and the Schwarschild-MOG black hole (MOG) \cite{Moffat}.

Black holes can accelerate particles to arbitrarily high energy if
the angular momentum of the particle is fine-tuned to some critical
value (see \cite{zak} and references therein). This phenomenon is
robust as it is founded on the basic properties of geodesics  around
a black hole \cite{Ac}. Studying the dynamics of a particle (either
massive or massless) around the gravitational source such as a black
hole (BH) is important because it is responsible for understanding
the geometrical structure of spacetime near the BH. Geodesics may
display a rich structure and convey a very reliable information to
understand the geometry of the BH. There are many types of geodesic
motion but the circular geodesics are specially important. The
exponential fade-out of a collapsing star's luminosity can be
explained by the circular geodesics as given in
\cite{Null1,Null2}. The motion of test particles helps to study the
gravitational fields of objects experimentally and to compare the
observations with the predictions about observable effects (light
deflection, gravitational time delay and perihelion shift).

In the surrounding of the BH, a magnetic field is generally present
\cite{new}, due to the presence of plasma in the vicinity of the BH.
The accretion disk or a charged gas cloud  is primarily responsible
for the magnetic field field \cite{1,2}. The magnetic field is
stronger in the vicinity of BH's event horizon however it does not
effect the geometry of the BH, but the motion of the charged
particle moving around a BH is effected \cite{p9,p9a}. The magnetic
coupling process is likely responsible for the stability of black
hole with its accretion disc \cite{3}. According to this process,
angular momentum and energy are transferred from the black hole to
its surrounding disc. The magnetic field plays an important role in
transferring sufficient energy to the surrounding particles for
escaping to spacial infinity \cite{4,5}. Other interesting processes
around BHs include evaporation and phantom energy accretion onto BHs
\cite{jamil}. In this article, we revisit the model of Zahrani et al
\cite{13} for a MOG black hole and explore the effects of modified
gravity. It involves the collision of a bounded particle with an
unbounded particle in the vicinity of the black hole. The main
interest lies in finding the conditions of escape of a bounded
particle after the collision.

The outline of the paper is as follows: In section II we develop the
basic equations and then derive an expression for escape velocity of
a neutral particle. In section III we discuss  the strength of
magnetic field, and the equations of motion of the charged particle
moving around weakly magnetized MOG and escape velocity for
particle also calculated. Force acting on the particle is studied in
section IV and geodesics of the particles moving around the MOG are
discuss in section V. Lyapunov exponent is explained in section VI.
In section VII and VIII, trajectories for effective potential and
escape velocity of the particle are presented respectively.
Conclusion is given in last section. We will study the motion of
particle in the equatorial plane to simplify the calculations.
Throughout this work we use the following metric signature
$(+,-,-,-)$ and assume $c=1$.

\section{ Neutral particle dynamics around MOG black hole}

Motion of particles around a central massive object under the affect
of a central force is a well-studied problem of classical mechanics
(or rather Newtonian mechanics). In particular we can think of the
following problem in the present context: consider a particle of
mass $m$ moving in a circular orbit around another object of mass
$M$ such that $M>>m$. Now for the particle to escape from the
gravitational field of $M$, particle's initial velocity must be more
than the escape velocity. The particle can gain the escape velocity
either  from an external force acting on it or by hitting (or
colliding) a test particle with the particle in circular orbit.
Since the collision leads to transfer of energy as well, the
particle will escape from the circular orbit if its energy after
collision is more than a critical energy or escape energy. If
however, the energy of the particle after collision is small than
the critical energy, the particle falls towards $M$.

The relativistic version of the above scenario was investigated by
Zahrani et al \cite{13}. They studied the motion of a charged
particle in the vicinity of a weakly magnetized Schwarzschild black
hole and focused on the bounded trajectory lying in the black hole
equatorial plane. For the charged particle to escape from the
innermost circular orbit, another particle (which is neutral and
coming from sufficiently far distance) hits the charged particle.
The authors obtained the corresponding conditions of escape velocity
and escape energy in the resulting process. They also predicted that
the motion of charged particle after collision will be chaotic due
to the presence of magnetic field and strong gravitational field.
Although the chances of collision between two particles around black
hole, in general are feeble. However the process itself is important
to describe the ejection of particles from the vicinity of black
holes. Later on Hussain et al \cite{saqib} investigated a similar
scenario for a slowly rotating Kerr black hole and discussed the
conditions of escape for the particle. Jamil et al \cite{jamil1}
investigated a similar scenario for a Schwarzschild black hole
surrounded by quintessence.

Recently, Moffat \cite{Moffat} obtained both static and non-static
black hole solutions in the scalar-tensor-vector modified gravity,
the theory which he himself proposed \cite{Moffat1}. The theory
fairly describes several astronomical and cosmological observations
such as galaxy rotation curves \cite{m1} and gravitational lensing
\cite{m2}. The modified gravitational field equations are given by
\cite{Moffat,Moffat1}:
\begin{equation}\label{1aa}
R_{\mu\nu}-\frac{1}{2}g_{\mu\nu}R =-8\pi G T^\varphi_{\mu\nu},
\end{equation}
where
$$ T^\varphi_{\mu\nu}=-\frac{1}{4\pi}\Big( B_\mu^{~\sigma} B_{\nu\sigma}-\frac{1}{4}g_{\mu\nu} B^{\sigma\beta}B_{\sigma\beta} \Big), $$
and $ B_{\mu\nu}=\varphi_{\nu,\mu}-\varphi_{\mu,\nu}, $ where $\varphi_\mu$
is a vector field with the source charge $Q=\sqrt{\alpha G_N} M$ (see details
below). The role of this vector field is to produce large scale
repulsive gravity which can cause accelerated cosmic expansion.
Further the vacuum field equations are
\begin{equation}
 B^{\mu\nu}_{~~;\nu}=0,~~~~ B_{[\mu\nu;\sigma]}=0
\end{equation}
where $;$ denotes the covariant derivative operation.

To solve the above system of equations, an ansatz for the static and
spherically symmetric solution is assumed of the form
\begin{eqnarray}\label{1}
ds^{2}&=& g(r)dt^{2}- g(r)^{-1}dr^2 -r^2(d\theta^2 +\sin^2\theta
d\phi^2).
\end{eqnarray}
The calculation yields \cite{Moffat}
\begin{eqnarray}\label{metric}
&&g(r)=1-2\frac{GM}{r}+\alpha\frac{ G_{N}G M^2}{r^2}, \nonumber\\&&
G=G_{N}(1+\alpha),\ \ Q=\kappa M,\ \  \kappa=\pm\sqrt{\alpha G_{N}},\ \ \alpha=\frac{G}{G_{N}}-1,
\end{eqnarray}
and therefore we have
\begin{equation}\label{m1}
  Q=\pm\sqrt{\alpha G_{N}}M.
\end{equation}
Note that $\alpha$ is a free parameter of the theory, hence it
yields a variable gravitational \textit{constant}. Here $M$ and $Q$
is respectively mass and electric charge of the black hole and
$G_{N}$ is the Newton's gravitational constant. In metric (\ref{1}),
positive value of $Q$ is chosen  to maintain repulsive gravitational
force as it is necessary to describe the stable star. (For a stable
star, the gravitational attraction should be balanced by the repulsive
gravity). This is a static, spherically symmetric, point particle
solution of a electrically charged body like Reissner-Nordstrom black hole \cite{RN,RN1}. For
$\alpha=0$, metric (\ref{1}) reduces to the Schwarzschild metric
(which is also the general relativistic limit). Like Kerr
\cite{Kerr} and Reissner Nordst$\ddot{o}$rm metrics it has two
horizons:
\begin{equation}\label{m2}
  r_{\pm}=G_{N}M\Big(1+\alpha\pm(1+\alpha)^{\frac{1}{2}}\Big).
\end{equation}
Eq. (\ref{m2}) corresponds to $g(r)=0$ and  will reduce to
Schwarzschild event horizon for $\alpha=0$. The
metric (\ref{1}) is invariant under time translation and rotation
around symmetry axis. Thus the Killing vectors equations are
\cite{p9b}:
\begin{equation}
\xi_{(t)}^{\mu}\partial_{\mu}=\partial_{t} , \qquad
\xi_{(\phi)}^{\mu}\partial_{\mu}=\partial_{\phi},
\end{equation}
which will give the constants of motion, where
$\xi_{(t)}^{\mu}=(1,~0,~0,~0)$ and
$\xi_{(\phi)}^{\mu}=(0,~0,~0,~1)$. The conserved quantities
corresponding  to these Killing vectors are the total energy (per
unit mass) $\mathcal{E}$ and azimuthal angular momentum $L_{z}$ (per
unit mass) of the moving particle at infinity. Motion of a neutral
particle moving in MOG background is described by the Lagrangian
density \cite{19},
\begin{equation}\label{m17}
  \mathcal{L}= \frac{1}{2}g_{\mu\nu}\dot{x}^{\mu}\dot{x}^{\nu}.
\end{equation}
From (\ref{metric}) and (\ref{m17}) we can say that $t$ and $\phi$
are the cyclic coordinates. There exists constants of motion
corresponding  to these cyclic coordinates i.e. total energy and
azimuthal angular momentum. We have calculated these integral of
motion by using the Euler-Lagrange equation
\begin{equation}\label{m18a}
  \frac{d}{d\tau}\Big(\frac{\partial\mathcal{L}}{\partial\dot{x}^{\mu}}\Big)-\frac{\partial\mathcal{L}}{\partial x^{\mu}}=0.
\end{equation}
Therefore, using equation (\ref{m18a}) for $t$ and $\phi$ we have
\begin{equation}\label{2}
 \frac{dt}{d\tau} =\dot{t}=\frac{\mathcal{E}}{g(r)},
\end{equation}
\begin{equation}\label{3}
  \frac{d\phi}{d\tau}=\dot{\phi}=-\frac{L_{z}}{r^2}.
\end{equation}
The over dot denotes the differentiation with respect to proper time
$\tau$  throughout the calculations. Considering the
planar motion of the particle i.e. for $\theta=\pi/2$ the
normalization condition $u^{\mu}u_{\mu}=1$, gives
\begin{eqnarray}\label{4}
\dot{r}^{2}&=&\mathcal{E}^2- U_\text{eff},
\nonumber\\
U_\text{eff}&=& \big(1-2\frac{GM}{r}+\alpha\frac{ G_{N}G M^2}{r^2}\big)\big(1 +\frac{L_z^2}{r^2}\big).
\end{eqnarray}
The extreme values of the effective potential correspond to
$\frac{dU_\text{eff}}{dr}=0$. It occurs at $r=6M$ for a
Schwarzschild black hole \cite{13}. The point where the ISCO exists
is the convolution point of the effective potential \cite{14}. In
the present case, the ISCO occurs at
\begin{eqnarray}\label{m12}
r_{o}&=& \frac{\alpha  G M^2 G_N+L^2}{3 G M}-\left(\sqrt[3]{2} \left(L (L-3 G M)+\alpha  G M^2 G_N\right) \left(L (3 G M+L)+\alpha  G M^2 G_N\right)\right)
\nonumber\\&&
\Bigg[3GM\Bigg(-2 \alpha ^3 G^3 M^6 G_N^3+3 G L^4 M^2 \left(9 G-2 \alpha  G_N\right)-2 L^6-3 \alpha  G^2 L^2 M^4 G_N \left(2 \alpha  G_N+9 G\right)
\nonumber\\&&
+3 \sqrt{3}\Big(G^3 L^2 M^4\bigg(-9 G L^6+108 G^3 L^4 M^2+\alpha  G_N\big(8 L^6-126 G^2 L^4 M^2
\nonumber\\&&
+\alpha  G M^2 G_N\big(24 L^4-9 G^2 L^2 M^2+8 \alpha  G M^2 G_N \left(\alpha  G M^2 G_N+3 L^2\right)\big)\big)\bigg)\Big)^{\frac{1}{2}}
\Bigg)^{\frac{1}{3}}\Bigg]^{-1}
\nonumber\\&&
-\frac{1}{3 \sqrt[3]{2} G M}
\Bigg(-2 \alpha ^3 G^3 M^6 G_N^3+3 G L^4 M^2 \left(9 G-2 \alpha  G_N\right)-2 L^6-3 \alpha  G^2 L^2 M^4 G_n \left(2 \alpha  G_N+9 G\right)
\nonumber\\&&
+3 \sqrt{3}\Big(G^3 L^2 M^4\bigg(-9 G L^6+108 G^3 L^4 M^2+\alpha  G_N\big(8 L^6-126 G^2 L^4 M^2
\nonumber\\&&
+\alpha  G M^2 G_N\big(24 L^4-9 G^2 L^2 M^2+8 \alpha  G M^2 G_N \left(\alpha  G M^2 G_N+3 L^2\right)\big)\big)\bigg)\Big)^{\frac{1}{2}}
\Bigg)^{\frac{1}{3}}
\end{eqnarray}
The critical energy and the azimuthal angular momentum of the
particle corresponding to ISCO are
\begin{equation}\label{m3}
\mathcal{E}_{o}=\frac{\left(\alpha  G M^2 G_N+r (r-2 G M)\right){}^2}{r^2 \left(2 \alpha  G M^2 G_N+r (r-3 G M)\right)},
\end{equation}
\begin{equation}\label{cir2}
\mathcal{L}_{zo}=\frac{\sqrt{G M r^3-\alpha  G M^2 r^2 G_N}}{\sqrt{2 \alpha  G M^2 G_N-3 G M r+r^2}}.
\end{equation}
We consider the case that an incoming  particle collides with the
orbiting particle at the ISCO, so that after collision it will move
within a new plane tilted with respect to previous equatorial plane.
However, to study the dynamics of a particle  it is convenient to
use the fact that if the initial position and the tangent vector of
the trajectory of the particle lies on a plane that contain the
center of the body, then the entire trajectory lies on this plane.
After collision, there are three possible cases depending on the
collision mechanism : (i) bound motion (ii) capture by BH (iii)
escape to infinity. For a little change in energy and angular
momentum, orbit of the particle alters very slightly. While  for
large changes it may escape to infinity or capture by BH depending
upon the nature of change in path. After collision, the particle
will no longer remain in the same equatorial plane, so further
discussion would be dealt with respect to new plane. But note that
due to spherical symmetry all equatorial planes are equivalent. Due
to collision particle should have new constants of motion
$\mathcal{E}$, $L^{2}$ and $L_{z}$. For simplification of our
problem we consider the case of collision when $(i)$ the azimuthal
angular momentum is invariant during collision $(ii)$ initial radial
velocity does not change. These conditions imply that only energy of
the particle will change hence its motion would be determined by
considering only the change in the energy. After collision particle
acquires an escape velocity $v$, in orthogonal direction of the
equatorial plane as explained in \cite{new}. The angular momentum
and energy of the particle after collision becomes (at the
equatorial plan $\theta=\frac{\pi}{2}$)
\begin{equation}\label{m10}
  L^{2}=r_{o}^{2}v^{2}+\frac{L_{zo}^{2}}{\sin^{2}\theta},
  \ \ L^{2}=r_{o}^{2}v^{2}+L_{zo}^{2}
\end{equation}
Here $v\equiv -r_{o}\dot{\theta}_{o}$ and $\dot{\theta}_{o}$ is the
initial polar angular velocity of the particle and $r_{o}$ is the
radius of ISCO. This velocity should be in orthogonal direction to
the equatorial plane \cite{punsly}.  Further
\begin{equation}\label{m5}
 \mathcal{E}=\sqrt
{(1-\frac{2GM}{r}+\frac{\alpha G_{N}G M^2}{r^2})v^2 + \mathcal{E}^2_o},
\end{equation}
where $\mathcal{E}_{o}$ is defined in Eq. (\ref{m3}). It is clear
that these values of angular momentum (\ref{m10}) and energy
(\ref{m5})are larger than their values (\ref{m3}) and (\ref{cir2})
before collision.
 Also Eq. (\ref{m5}) shows that as
$r\rightarrow\infty$,
$\mathcal{E}\rightarrow\mathcal{E}_{o}\rightarrow 1$. So for
$\mathcal{E}\equiv \mathcal{E} \geq1$ particle will have
unbound motion. In other words the particle can not escape to
infinity if $\mathcal{E}<1$. Hence  the necessary condition  for a particle to escape
to infinity after collision is $\mathcal{E}\geq1$ or
\begin{equation}\label{m4}
v\geq \frac{\sqrt{r \left(2 G M \left(L^2+r^2\right)-L^2 r\right)-\alpha  G M^2 G_N \left(L^2+r^2\right)}}{\sqrt{r^2 \left(\alpha  G M^2 G_N+r (r-2 G M)\right)}}.
\end{equation}
The last expression for velocity $v$ is obtained by solving equation
$(\ref{m5})$ at $\mathcal{E}_\text{new}=1$.

\section{Motion of a Charged Particle Around a Weakly  Magnetized SMOG Black Hole}

\subsection{Magnetic Field in the Vicinity of BH}

The magnetic coupling (MC) process is responsible for  attraction of
black hole with its accretion disc \cite{p9, 3, r1, r2, r3}.
According to this process angular momentum and energy are
transferred from a black hole to its surrounding disc. The process
of MC provides the relation between the strength of magnetic field
at the black hole horizon and its mass $M$ and the rate of accretion
$\dot{M}$ \cite{M}. This relation is as follows:
\begin{equation}\label{m15}
  \mathcal{B}_{h}=\frac{1}{r_{h}}\sqrt{2m_{p}\dot{M}c}.
\end{equation}
Here black hole horizon $r_{h}$ is given by (\ref{m2}) and $m_{p}$
is the magnetization parameter which indicates the relative power of
process of MC with respect to disc accretion. If  disc accretion
dominant over MC process then $m_{p}<1$ and if MC process is
dominant over disc accretion then $m_{p}>1$ while $m_{p}=1$
correspond to equipartition of these two processes. The magnetic
field expression is given by \cite{M}
\begin{equation}\label{m16}
\mathcal{B}_{h}=\Big(v_{b}f^{2}(\alpha,m_{p})\Big)^{\frac{1}{2}}
\times10^{7.35+.45\theta}.
\end{equation}
Magnetic field strength at the horizon of MOG-BH is $4.93\times10^{8}$ Gauss
for $m_{p}=1$, $\alpha=0.1$, $\theta=0.5$ and $v_{b}=300$.


Similar effects of  particle collision with high center of mass
energy in the vicinity of black hole can also possible if the black
hole is non rotating provided there exists a magnetic field in its
surrounding. Since, there exists theoretical and experimental
evidence that such a magnetic field should exist in the surrounding
of black hole \cite{5}-\cite{black}. We assume that this magnetic
field is weak and its energy and angular momentum does not effect
the background geometry of the black hole. The above mentioned
condition satisfy for a black hole of mass $M$ if magnetic field
strength holds the condition \cite{mag}
\begin{equation}\label{m14}
\mathcal{B}<<\mathcal{B}_{max}=\frac{1}{G^{\frac{3}{2}}M_{\bigodot}}\Big(\frac{M_{\bigodot}}{M}\Big)\sim10^{19}\frac{M_{\bigodot}}{M}.
\end{equation}
These kind of black holes are called weakly magnetized.

\subsection{Magnetic Field Calculation}

In this section we explore how does the presence of a magnetic field
in the BH exterior stimulate the motion of a charged particle.
Firstly, we will calculate the magnetic field in the vicinity of
black hole by following the procedure as given by \cite{saqib, 13}

\begin{figure}[!ht]
\centering
\includegraphics[width=6cm]{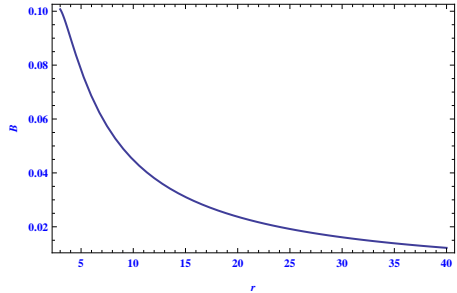}
\caption{In this figure we have plotted the magnetic field $B^{\theta}$ vs
$r$ for $\alpha=0.2$.} \label{fMOG17}
\end{figure}

The Killing vector equation is \cite{10}
\begin{equation}\label{34}
  \Box\xi^{\mu}=0,
\end{equation}
where $\xi^{\mu}$ is a Killing vector. Eq. (\ref{34}) corresponds to
the Maxwell equation for 4-potential $A^{\mu}$ in the Lorentz gauge
$A^{\mu}_{\ \ ;\mu}=0$. Defining $A^{\mu}$ as \cite{6a,6}
\begin{equation}\label{m7}
  A^{\mu}=\Big(\frac{\alpha GM}{r},0,0,\frac{\mathcal{B}}{2}\Big).
\end{equation}
Here magnetic field is defined as \cite{13}
\begin{equation}\label{5}
  \mathcal{B}^{\mu}=-\frac{1}{2}e^{\mu\nu\lambda\sigma}F_{\lambda\sigma}~u_{\nu},
\end{equation}
where
\begin{equation}\label{6}
  e^{\mu\nu\lambda\sigma}=\frac{\epsilon^{\mu\nu\lambda\sigma}}{\sqrt{-g}},\ \
  \epsilon_{0123}=1,\ \ g=det(g_{\mu\nu}).
\end{equation}
Here $\epsilon^{\mu\nu\lambda\sigma}$ is the Levi Civita symbol, and
$ F_{\mu\nu}$ is the Maxwell tensor, defined as
\begin{equation}\label{7}
  F_{\mu\nu}=A_{\nu,\mu}-A_{\mu,\nu}=A_{\nu;\mu}-A_{\mu;\nu}.
\end{equation}
For a local observer at rest, the component of four velocity are
\begin{equation}\label{m18}
  u^{\mu}=(u^{t},0,0,0)=\Big(g(r)^{\frac{-1}{2}},0,0,0\Big).
\end{equation}
Here we assume $u^t>0,$ to signify the `forward-in-time' condition.
By equations (\ref{m7}-\ref{m18}) the magnetic field $4$-vector
becomes
\begin{eqnarray}\label{1a}
  \mathcal{B}^{\mu}&=&\Bigg(0,\mathcal{B}g(r)^{\frac{1}{2}}\cos\theta,
  -\frac{\mathcal{B}g(r)^{\frac{1}{2}}}{r}\sin\theta
 ,0\Bigg).
\end{eqnarray}
The magnetic field is along with the $z$-axis at spatial infinity,
and we assume that it is directed upward \cite{35}. At equatorial
plane only $B^{\theta}$ component will survive. In figure
\ref{fMOG17} we have plotted $B^{\theta}$ as a function of $r$. It
is decreasing initially and then become almost constant for large
$r$ (away from the black hole). So, it is homogeneous at
$r\rightarrow\infty$.

\subsection{Dynamical Equations}

The Lagrangian of the moving particle of rest mass $m$  and electric
charge $q$ in the vicinity of MOG-BH is
\begin{equation}\label{10}
  \mathcal{L}=\frac{1}{2}g_{\mu\nu}\dot{x}^{\mu}\dot{x}^{\nu}+\frac{q}{m}A_{\mu}\dot{x}^{\mu}.
\end{equation}
Using Euler-Lagrange equation (\ref{m18}) for $t$ and $\phi$ we get
 \begin{equation}\label{11}
 g(r)\Big(\dot{t}+\frac{\epsilon\alpha GM}{r}\Big)=\mathcal{E},
\end{equation}
Here, $\epsilon=\frac{q}{m}$ is the specific charge of a particle and
\begin{equation}\label{11a}
  \dot{\phi}=-\frac{L_{z}}{r^{2}\sin^{2}\theta}+B,
\end{equation}
where
\begin{equation}\label{12}
  B\equiv\frac{\epsilon\mathcal{B}}{2}.
\end{equation}
Using the  normalization condition $u^{\mu} u_{\mu}=1$, we obtain
\begin{equation}\label{15a}
1=g(r)\dot{t}^{2}-\frac{1}{g(r)}\dot{r}^2 -r^2 \dot{\theta}^2 -r^{2}\sin^{2}\theta\dot{\phi}^{2}.
\end{equation}
By using equations (\ref{11}) and (\ref{11a}) in (\ref{15a}) and choosing $\theta=\frac{\pi}{2}$
we have
 \begin{equation}\label{15b}
 \dot{r}^{2}+U_\text{eff}=\mathcal{E}^{2},
\end{equation}
and then effective potential is
 \begin{equation}\label{m8}
\mathcal{E}_{\pm}= U_\text{eff}{\pm}=\frac{\epsilon GM}{r}\pm \sqrt{g(r)\Big[1+r^2(\frac{L_{z}}{r^2 }-B)^2\Big]}.
\end{equation}
According to equation (\ref{m10}), after collision $L_{z}\rightarrow L$.
Hence the effective potential reduces to
 \begin{equation}\label{m8}
\mathcal{E}_{\pm}= U_\text{eff}{\pm}=\frac{\epsilon\alpha GM}{r}\pm \sqrt{g(r)\Big[1+r^2(\frac{L}{r^2 }-B)^2\Big]}.
\end{equation}
Putting the value of $L$ from equation (\ref{m10}) and $\mathcal{E}_{\pm}=1$ in equation (\ref{m8})
and then solving for $v$ we get
\begin{eqnarray}\label{m11}
v&=& \frac{1}{r^4 \left(\alpha  G M^2 G_N+r (r-2 G M)\right)}
\Bigg[G M r^2 \left(B r^2-L\right) \left(\alpha  M G_N-2 r\right)+B r^6-L r^4
\nonumber\\&&
\pm\sqrt{G M r^6 \left(\alpha  G M^2 G_N+r (r-2 G M)\right) \left(\alpha ^2 G M \epsilon ^2-\alpha  M G_N+r (2-2 \alpha  \epsilon )\right)}
\Bigg].
\end{eqnarray}
For a particle to escape from the black hole's vicinity, its velocity
should be greater or equal to $v$.
\par
\begin{figure}[!ht]
\centering
\includegraphics[width=10cm]{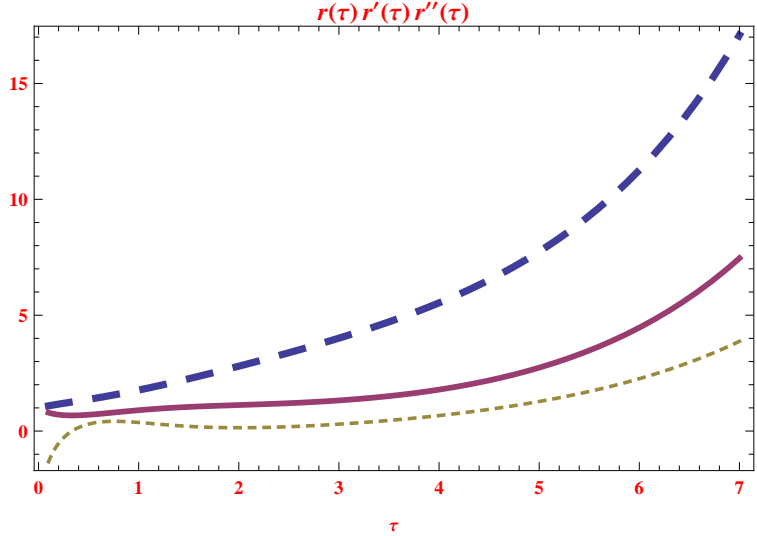}
\caption{Interpolating function $r(\tau)$ as the solution of
Eq. $(\ref{14a})$ for $L_{z}=2$, $B=0.3$, and $\alpha=0.2$. Here
the large bold dashed curve represents $r{\tau}$, solid curve is for ${\dot r(\tau)}$
and Short dashed curve correspond to ${\ddot r(\tau)}$.} \label{fMOG14}
\end{figure}

A charged particle moving in an external electromagnetic field
$F_{\mu\nu}$ obeys the equation of motion:
\begin{equation}\label{geodesic}
 \ddot {x}^{\mu}+\Gamma^{\mu}_{\nu \sigma} \dot{x}^{\nu} \dot{x}^{\sigma}=\frac{q}{m}F^{\mu}_{\alpha} \dot{x}^{\alpha}.
 \end{equation}
The dynamical equations for $\theta$ and $r$ respectively are
\begin{eqnarray}\label{13}
  \ddot{\theta}&=&\frac{-2}{r}\dot{r}\dot{\theta} +\frac{L^2_{z}\cos\theta}{r^4 \sin^3\theta}-B^2 \sin\theta
  \cos\theta,\\
  \ddot{r}&=&\gamma -\dot{\theta}^2 \Big(\gamma r^2 -r g(r)\Big)-\frac{L^2}{r^2
  \sin^2\theta}\big(\gamma-\frac{g(r)}{r}\Big)
  \nonumber\\&&\label{14}
  -B^2 \sin^2\theta\Big(-3r g(r)+r^2\gamma\Big)+B L\Big(2\gamma-\frac{4g(r)}{r}\Big),
\end{eqnarray}
where $\gamma=\frac{M}{r^2}-\frac{\alpha G M^2}{r^3}$.
For $\theta=\frac{\pi}{2}$ the equation (\ref{13}) is satisfied obviously
and equation (\ref{14}) becomes
 \begin{eqnarray}\label{14a}
  \ddot{r}&=&\gamma -\frac{L^2}{r^2}\Big(\gamma-\frac{g(r)}{r}\Big)-B^2 (-3r g(r)
  +r^2\gamma)+B L\Big(2\gamma-\frac{4g(r)}{r}\Big).
\end{eqnarray}
We have solved equation (\ref{14a}) numerically by using the built
in command of Mathematica NDSolve for $\alpha=0.2$, $B=0.3$ and
$L_{z}=2$ and plotted the solution in figure \ref{fMOG14}. In this
figure the upper curve represents $r(\tau)$, middle is for $\dot{r}$
(radial velocity) and the lower one is for $\ddot{r}$. $\dot{r}$ is
the radial velocity which represents the escape behavior as it is
increasing with the increase of $r$, and $\ddot{r}$ is the radial
acceleration which is analogous to the force exerted on the particle
in the radial direction after the collision.

\section{Force on a charged particle in the vicinity of MOG-BH}

As we have already calculated the effective potential for MOG-BH,
we can also compute the effective force on a neutral and a charged
particle by \cite{Fernendo},
\begin{equation}\label{m13}
F=-\frac{1}{2}\frac{dU_\text{eff}}{dr}.
\end{equation}
\begin{equation}\label{ma}
F=\frac{-G_N M \left(3 L^2+r^2\right)+L^2 r}{r^4}+\frac{\alpha  G_N M \left((\alpha +1) G_N M \left(2 L^2+r^2\right)-r \left(3L^2+r^2\right)\right)}{r^5}.
\end{equation}
It can be seen from  equation (\ref{ma}) that the force due to
scalar tensor-vector gravity is repulsive if
$(\alpha +1) G_N M \left(2 L^2+r^2\right)>-r \left(3 L^2+r^2\right)$.
\begin{equation}\label{ma1}
F=\frac{L^2 r-G M \left(3 L^2+r^2\right)}{r^4}+\frac{Q^2 \left(2 L^2+r^2\right)}{r^5}
\end{equation}
Equation (\ref{ma1}) represents the effective force for RN-BH and the force
due to charge of BH is repulsive with out any condition.

\begin{figure}[!ht]
\centering
\includegraphics[width=10cm]{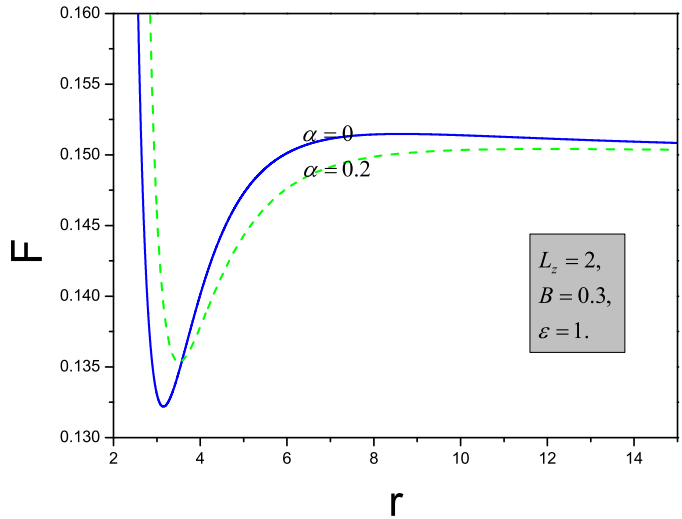}
\caption{Here we have plotted effective force as function of $r$
for different values of $\alpha$.} \label{fMOG15}
\end{figure}

We are studying the dynamics of a neutral and a charged particle
in the surrounding of MOG-BH where the scalar tensor-vector field produces a repulsive
gravitational force which prevents a particle to fall into singularity
\cite{Moffat1}.
In figure \ref{fMOG15} we are comparing the effective force on a particle in the vicinity
of MOG-BH with Schwarzschild black hole. We deduce from this figure that the
repulsion to reach the singularity is more for $\alpha=0.2$ as compared to $\alpha=0$.

\begin{figure}[!ht]
\centering
\includegraphics[width=10cm]{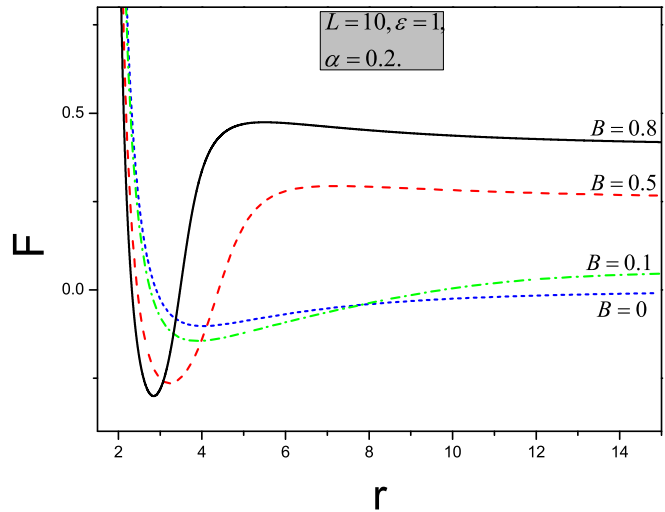}
\caption{Here we have plotted effective force against $r$
for different values of magnetic field $B$.} \label{fMOG16}
\end{figure}

To study the behavior of force against magnetic field $B$ we have
plotted the force $F$ as a function of $r$ for different value of
$B$ in figure \ref{fMOG16}. It can be seen from this figure, due to large magnetic
field strength $B$, force on the particle is more attractive as
compared to small $B$.

\section{Comparison of Geodesics in the Vicinity of MOG-BH vs SBH}

\begin{figure}[!ht]
\centering
\includegraphics[width=10cm]{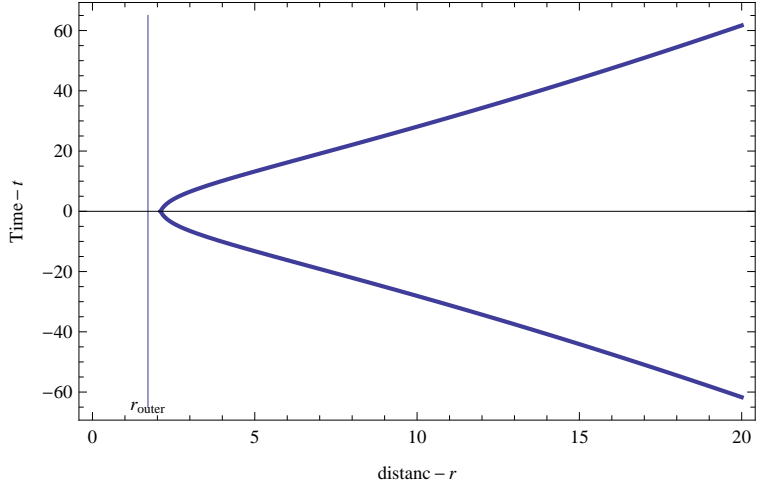}
\caption{Here we have plotted the geodesic equations $\frac{dt}{dr}$ (\ref{g1}) against $r$ for
 $\mathcal{E}=1$, $L_z=2$, $M=1$.} \label{fMOG18}
\end{figure}

\begin{figure}[!ht]
\centering
\includegraphics[width=10cm]{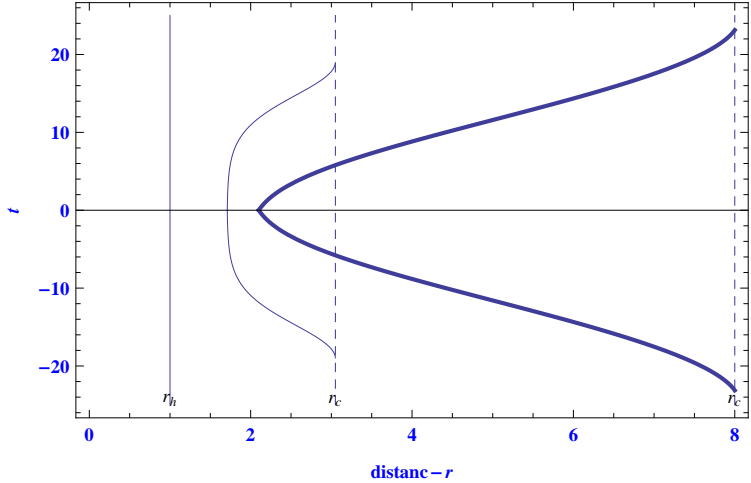}
\caption{Here we have plotted the geodesic equations $\frac{dt}{dr}$ (\ref{geo1}) against $r$ for
 $\mathcal{E}=1$, $L_z=2$, $M=1$. } \label{fMOG13}
\end{figure}

\textbf{Geodesics of a Neutral Particle Moving Around a Schwarzschild BH:}
\\
Geodesics of a particle approaching toward or  away from BH could be obtained by
using Eqs. (\ref{2}) and (\ref{4}) together. We have:
\begin{equation}\label{g1}
\frac{dt}{dr}=\pm \frac{\mathcal{E}}{g(r)}\frac{1}{\sqrt{\mathcal{E}^{2}-U_\text{eff}}} ,
\end{equation}
Here $U_\text{eff}$ correspond to a neutral particle given by (\ref{4}),
where positive root gives the path of the  particle out-going from
the BH, and negative root gives the path of an ingoing particle. Let
us consider the particle which is coming from infinity, initially at
rest, approaches the BH, so setting $\mathcal{E}=1$, $B=0$, $\alpha=0$ and $L_{z}=2$ in Eq.
(\ref{g1}) we plot the geodesics in $(r, t)$ coordinates, see figure
\ref{fMOG18}.
\\
{\bf Geodesics of a Charged Particle Moving Around a MOG-BH:}
\\
Geodesics of the particles approaching the MOG-BH could be
obtained by using Eqs. (\ref{11}) and (\ref{15b}) together we obtain:
\begin{equation}\label{geo1}
\frac{dt}{dr}=\pm \sqrt{\frac{\mathcal{E}g(r)^{-1}-\frac{\epsilon\alpha GM}{r}}{\mathcal{E}^{2}-U_\text{eff}}} ,
\end{equation}
Here $U_\text{eff}$ corresponds to a charged particle given by
equation (\ref{m8}) where positive and negative signs give the path
of the outgoing and ingoing particle respectively. Setting
$\mathcal{E}=1$, $L_z=2$, $M=1$, in Eq. (\ref{geo1}) we get the
geodesics which are bounded by the boundaries $r=r_c$ and the outer
horizon of the BH, plotted in figure \ref{fMOG13}. In this figure,
for thick curve we consider parameter $\alpha=0.1$, magnetic field
strength $B=0.1$,  and charge of a particle $\epsilon=0.5$ and thin
curve correspond to $\epsilon=1$, $B=0.5$ and $\alpha=0.5$. Figure
\ref{fMOG13} shows that if the strength of magnetic field $B$ is
higher, then charged particle can reach arbitrary close to black hole as
compared to smaller value of $B$.

\section{Stability of  Orbits}

\begin{figure}[!ht]
\centering
\includegraphics[width=10cm]{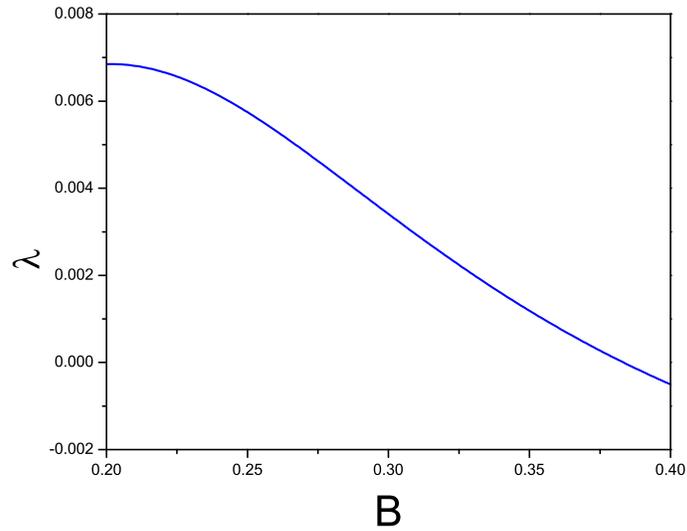}
\caption{In this figure we have plotted  the Lyapunov exponent as a function of magnetic
field $B$ for $L=6$, $\alpha=0.5$, $\mathcal{E}=1$, and $\epsilon=1$.} \label{fMOG11}
\end{figure}

\begin{figure}[!ht]
\centering
\includegraphics[width=10cm]{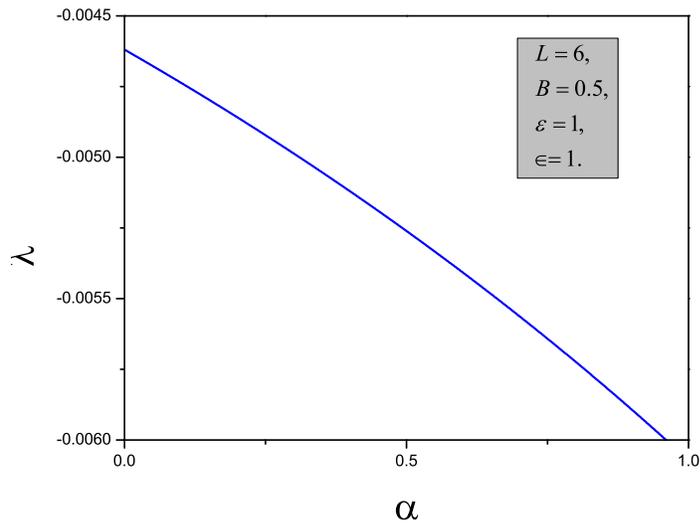}
\caption{Plotting of the Lyapunov exponent as a function of parameter
$\alpha$.} \label{fMOG12}
\end{figure}

Lyapunov exponents are the measurements of the rate of convergence
or divergence of the nearby trajectories in the phase space. It is
highly sensitive to the initial conditions. Its positive value is the
indication of divergence among the nearby trajectories. Therefore we
can check the stability of orbits by the Lyapunov exponent $\lambda$
\cite{cardoso}. It is given by
\begin{equation}\label{1.6}
\lambda=\sqrt{\frac{-U''_\text{eff}(r_o)}{2\dot{t}(r_o)^{2}}},
\end{equation}
where $r_o$ is the ISCO of the particle moving around BH.
In  \cite{cardoso} a critical component for the instability
is defined as
\begin{equation}\label{m12}
\gamma=\frac{T_{\lambda}}{T}, \ \ \ \ T_{\lambda}=\frac{1}{\lambda}
\end{equation}
Here $T_{\lambda}$ is the instability time scale.
Time period for a circular orbits can be obtained from equation
(\ref{3}) in the absence of magnetic field as
\begin{equation}\label{m12}
T=\mid\frac{2\pi r_{o}^{2}}{L_{z}}\mid,
\end{equation}
and in the presence of magnetic field, form equation (\ref{11a}) we have
 \begin{equation}\label{m12}
T=\mid\frac{2\pi r_{o}^{2}}{-L_{z}+Br_{o}^{2}}\mid
\end{equation}
Here $r_{o}$ is the radius of circular orbit.
We have plotted Lyapunov exponent
as a function of magnetic field $B$ in figure \ref{fMOG11}. It shows that the larger the
magnetic field strength the lesser will be $\lambda$.
From figure \ref{fMOG11} and equation (\ref{m12}) one can say that
the instability of circular orbits is more for a Schwarzschild black
hole in comparison with the black hole immersed in a magnetic field.
In figure \ref{fMOG12} we have also plotted $\lambda$ against $\alpha$
it decreases by the increase of $\alpha$. Hence, the stability of circular
orbits is less for the Schwarzschild black hole as compared to the black hole
 with non zero $\alpha$.
\pagebreak
\section{Behavior of Effective Potential}
Effective potential largely depends on $g_{00}$ component of metric.
Therefore, before discussing the behaviour of effective potential we will compare the $g_{00}$ component
of MOG-BH metric with RN-BH metric as it look similar. For RN-BH metric,
\begin{equation}\label{mg1}
  g_{00}=\Big(1-\frac{2M}{r}+\frac{Q^2}{r^2}\Big),
\end{equation}
for MOG-BH from equation (\ref{metric}) we have
\begin{equation}\label{mg2}
  g_{00}=\Bigg(1-\frac{2M}{r}-2\sqrt{\alpha}\frac{Q}{r}+\sqrt{\alpha}(1+\alpha)\frac{Q^2}{r^2}\Bigg).
\end{equation}
One can see that equation (\ref{mg1}) contain only $Q^2$ term while equation $(\ref{mg2})$ contain $Q$ and $Q^2$.
This difference leads to large change between the behaviour of effective potential of these two black holes.
The $g_{00}$ component of MOG-BH metric also contain a parameter $\alpha$ which will create a main difference
 between the behaviour of potential.

\begin{figure}[!ht]
\centering
\includegraphics[width=10cm]{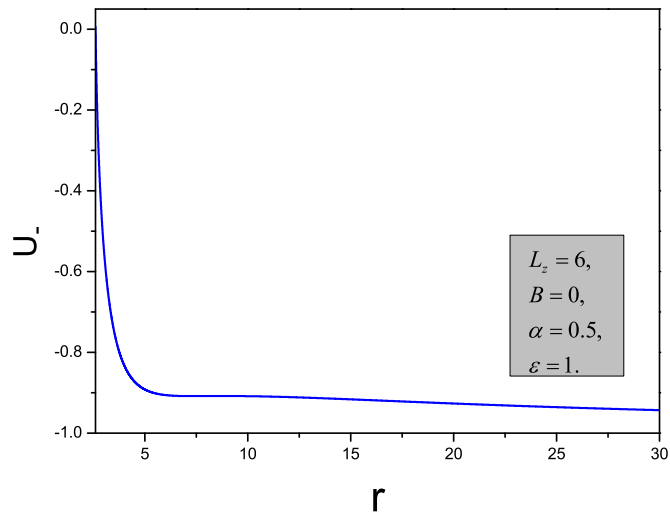}
\caption{In this figure we have plotted the effective potential $U_{+}$
against radial coordinate $r$. } \label{fMOG6}
\end{figure}

\begin{figure}[!ht]
\centering
\includegraphics[width=10cm]{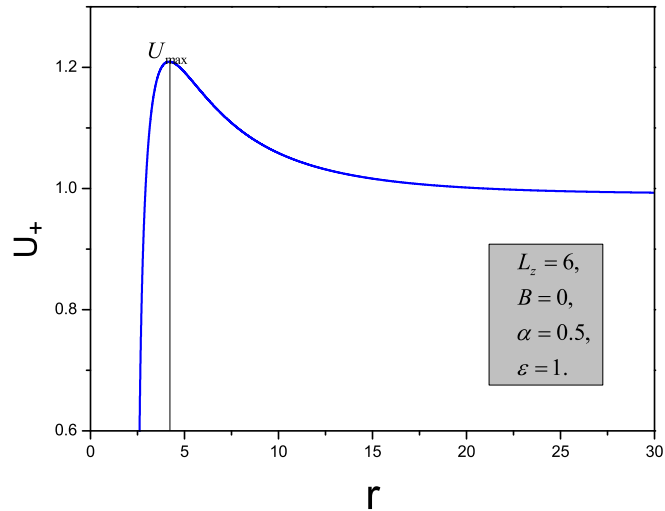}
\caption{In this figure we have plotted the effective potential $U_{+}$ as a
function  of $r$. } \label{fMOG7}
\end{figure}

In this section we plot the effective potential and graphically explain
the conditions on the energy of the particle required for escape to infinity or for bound motion around MOG-BH.
In figure \ref{fMOG6} and figure \ref{fMOG7} we have plotted $U_{-}\text{eff}$ and $U_{+}\text{eff}$
respectively correspond to equation (\ref{m8}). Here we will discuss the long distance
and short distance behaviour of equation (\ref{m8}). One can see that for small value of $r$
the square root term  will dominate but for large $r$ the term which is proportional to $\frac{1}{r}$
will dominate. For $U_{+}$ we get some maximum value which we have mentioned in figure \ref{fMOG7} as $U_{max}$.
For a particle to fall into the black hole its energy should be greater then  $U_{max}$ otherwise
it will bounce back to infinity or to some stable orbit. In case of $U_{-}$, particle can never cross the
barrier as shown in  the figure $\ref{fMOG6}$. Hence it cannot fall into the singularity.

\begin{figure}[!ht]
\centering
\includegraphics[width=12cm]{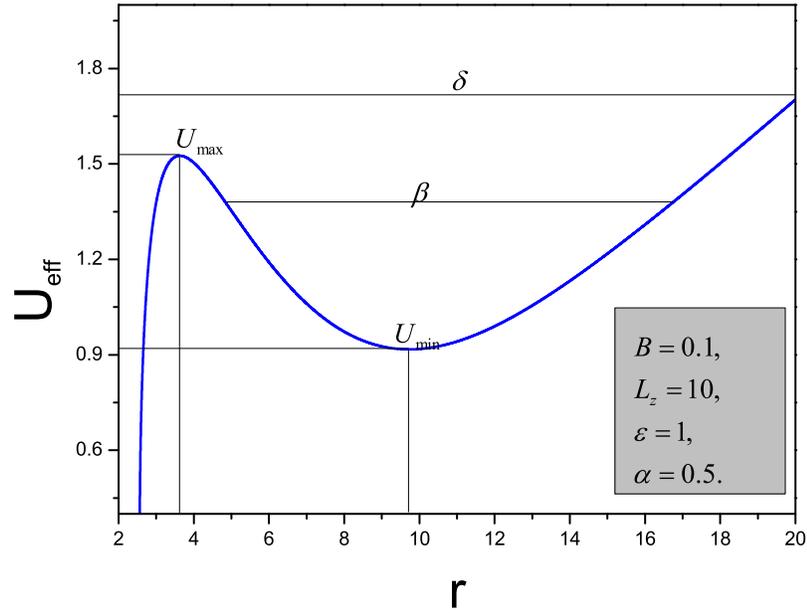}
\caption{In this figure different regions of effective potential which
correspond to escape and bound motion of the particle are shown. Here
 $\beta$ correspond to stable orbits for
$b=0.5$.}
\label{fMOG3}
\end{figure}

In figure \ref{fMOG3} different regions of effective potential which correspond to escape
and bound motion of the particle are shown. In figure \ref{fMOG3},
$\beta$  corresponds to region of stable orbits.
If the particle has energy equal or greater then $\delta$,
then it will fall into the black hole. $U_{max}$ and $U_{min}$
correspond to unstable and ISCO orbits respectively. If the energy of the
the particle is equal or less than $\beta$ then it will reside in one
of the stable orbits. Particle having energy greater than $\beta$ but less or
 equal to $\delta$ can have two possibilities: either it will escape
 to infinity or captured by the black hole. The
local minima might be related to the change in behavior of the
effective potential corresponds to BH (black brane transition) as
observed in \cite{Lyapunov1}.

\begin{figure}[!ht]
\centering
\includegraphics[width=12cm]{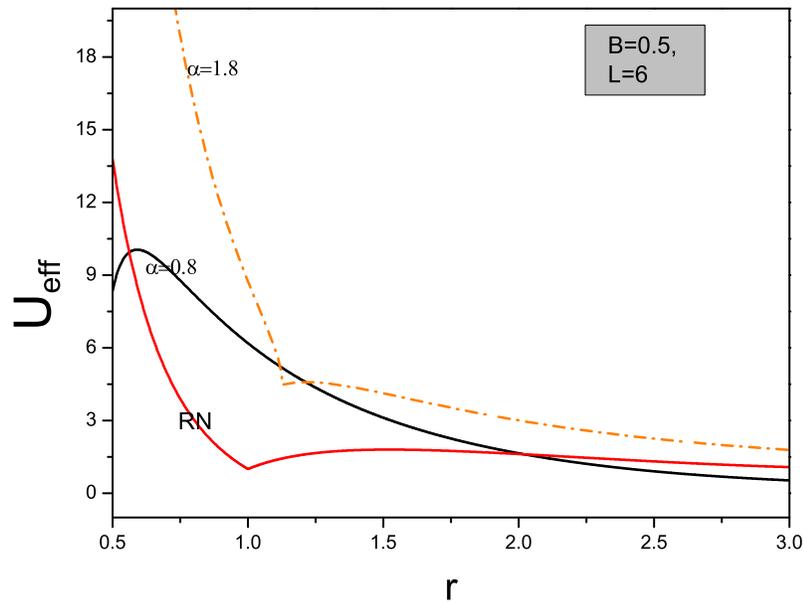}
\caption{Comparison of effective potential $U_\text{eff}$ as a function of $r$.
In this figure we are comparing the effective potential of
RN-BH and MOG-BH.} \label{fMOG1}
\end{figure}

In figure \ref{fMOG1} we are comparing the effective potential of Schwarzschild-BH,
RN-BH and MOG-BH. It can be seen from this figure that for large $r$ all the
potentials approach to $1$. Hence it can be concluded that the minimum energy
required for a particle to escape from black hole vicinity is $\mathcal{E}=1$ as
we have mentioned before.
Further the maxima for the effective potential
of RN-BH is greater in comparison with the maxima of effective potential of Schwarzschild-BH and MOG-BH.
Particle will be captured if it has energy greater than this maxima otherwise it
will move back to infinity or may be reside in some stable orbit.
Therefore, we can say that the possibility for a particle to escape
from the vicinity of a black hole or to reside in some stable orbit
 is high in case of MOG-BH with $\alpha=1.8$ in comparison
to RN-BH and MOG-BH with $\alpha=0.8$. Particle can reach to the singularity
depending upon its energy, but it cannot reach  the singularity for $\alpha=1.8$
as shown by the curve, in figure \ref{fMOG1}. Note that $\alpha>1$, corresponds to a naked singularity.

\begin{figure}[!ht]
\centering
\includegraphics[width=12cm]{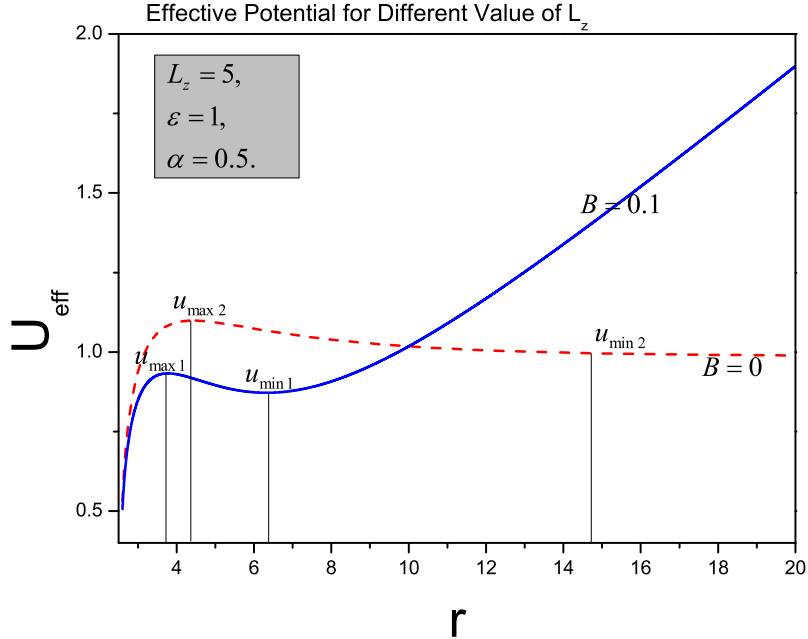}
\caption{Behavior of effective potentials with and without magnetic
field versus $r$ (a comparison).} \label{fMOG4}
\end{figure}

In figure \ref{fMOG4} we are comparing the effective potentials in the
presence with the absence of magnetic field.
In this figure $u_{max1}$ and $u_{max2}$ correspond to unstable orbits
while $u_{min1}$ and $u_{min2}$ is refers to ISCOs. One can
notice that in the presence of magnetic field, minima  of the effective potential is
shifted towards the horizon and width of stable region is also increased
in comparison with the case when the magnetic is absent. This is in agreement
with \cite{6, saqib}.
Therefore we can say that magnetic field act as to increase the
stability of the orbits.

\begin{figure}[!ht]
\centering
\includegraphics[width=12cm]{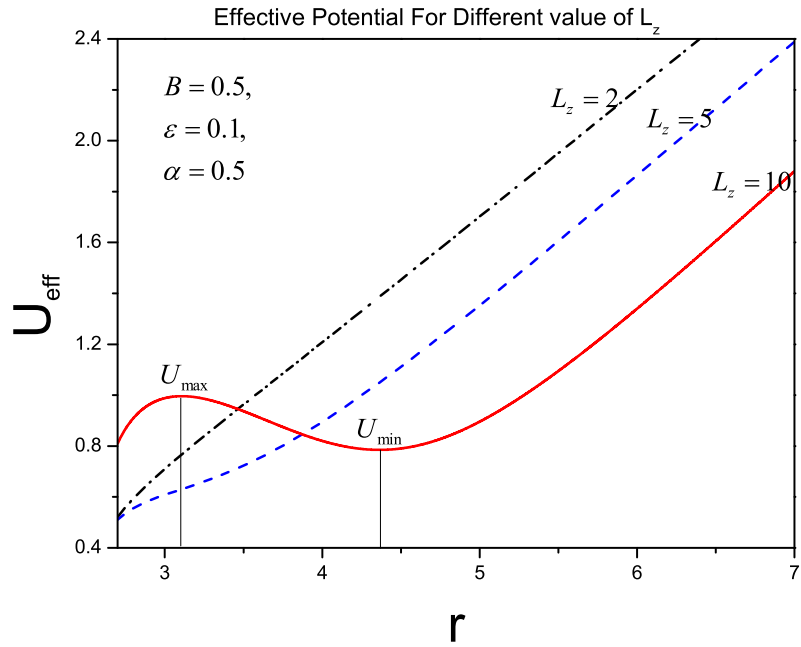}
\caption{Comparison of effective potentials with respect to $L_{z}$ as a function of $r$.}
\label{fMOG2}
\end{figure}

In figure \ref{fMOG2} we have plotted the effective potential as a function of
$r$ for different value of angular momentum $L_{z}$. One can see that for large value of
$L_{z}$, the effective potential has local minima and maxima which correspond to
stable and unstable circular orbits respectively. Hence we can say that the particle with greater value
of $L_{z}$ would have more possibility to reside in the stable orbits in comparison
with lesser value of $L_{z}$.

\begin{figure}[!ht]
\centering
\includegraphics[width=12cm]{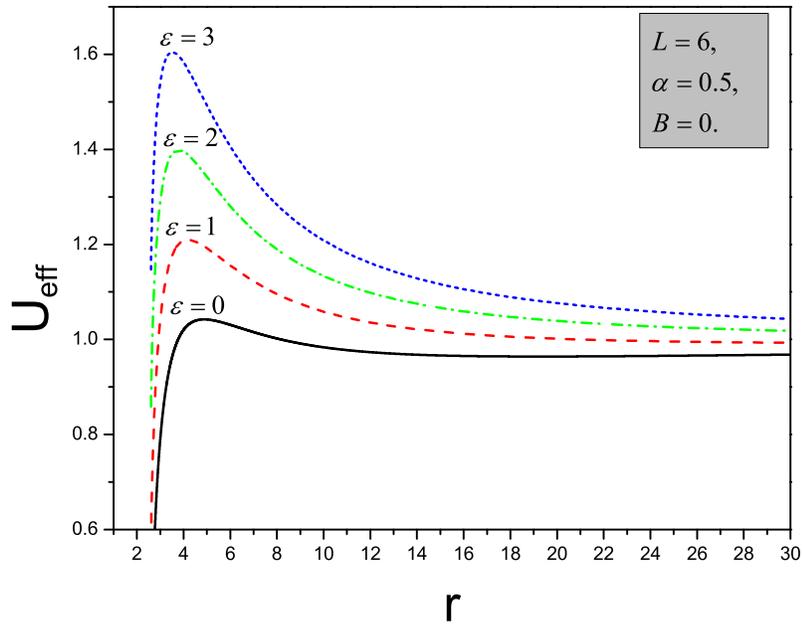}
\caption{Here we have plotted effective potential as function of $r$
for different values of $\epsilon$.} \label{fMOG10}
\end{figure}

To study the behaviour of effective potential against $\epsilon$,
we have plotted effective potential as a function of radial coordinate $r$
for different values of $\epsilon$ in figure \ref{fMOG10}.
It can be seen that larger  the value of $\epsilon$
greater will be the maxima of potential. Therefore, large value of $\epsilon$
would correspond to easy escape and vice versa.
It can be seen from the detailed analysis of the effective potential and the black hole physics that the
scalar tensor-vector modified gravity differs from the Einstein's theory of gravity a lot
at shorter distance and it becomes similar at long distance.
\section{Trajectories of Escape Velocity}

For the angular variable we have,
\begin{equation}\label{A}
  \frac{d\phi}{d\tau}=-\frac{L}{r^{2}}+B.
\end{equation}
If the left hand side of equation (\ref{A}) is negative then the Lorentz force on the particle
 is attractive \cite{11}. The motion of the charged particle is in clockwise direction.
 Lorentz force is repulsive if left hand side of (\ref{A}) is positive. We are not going to
the detail here because it is already discussed in \cite{11,14}. Our concern is only about the action of magnetic field on the charged particle. Therefore, magnetic field may deform the oscillatory motion,
so, the greater the strength of magnetic field, the larger will be the deformation of the orbit. Hence we can conclude that larger
the strength of magnetic field, easy for a particle to escape from the ISCO.

\begin{figure}[!ht]
\centering
\includegraphics[width=10cm]{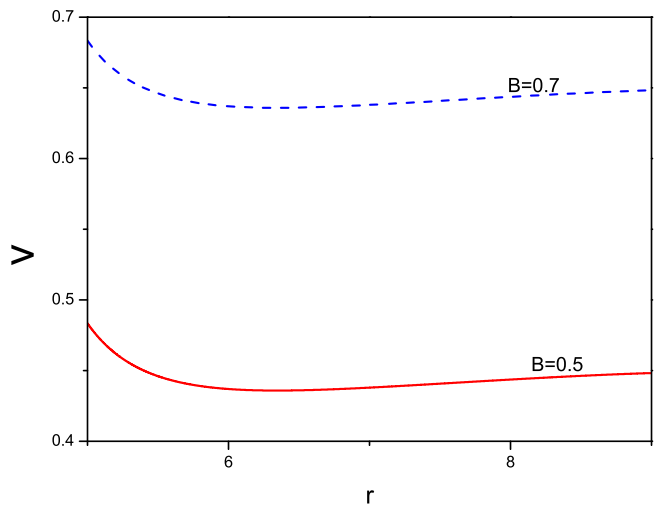}
\caption{Escape velocity ($v$) as a function of $r$ for
different values of magnetic field $B$. } \label{fMOG8}
\end{figure}

We have also plotted in figure \ref{fMOG8} escape velocity as a function of radial coordinate for different values of magnetic field parameter $b$. It can be seen
that the escape velocity of the particle increases as the magnetic
field strength increases but it becomes almost constant just like magnetic field, away from
the BH . As the magnetic field is strong near the BH, therefore, we can conclude that
presence of magnetic field will provide more energy to particle, so that it might easily escape from the vicinity of
BH. These conclusions are consistent with \cite{1,2}.

\begin{figure}[!ht]
\centering
\includegraphics[width=12cm]{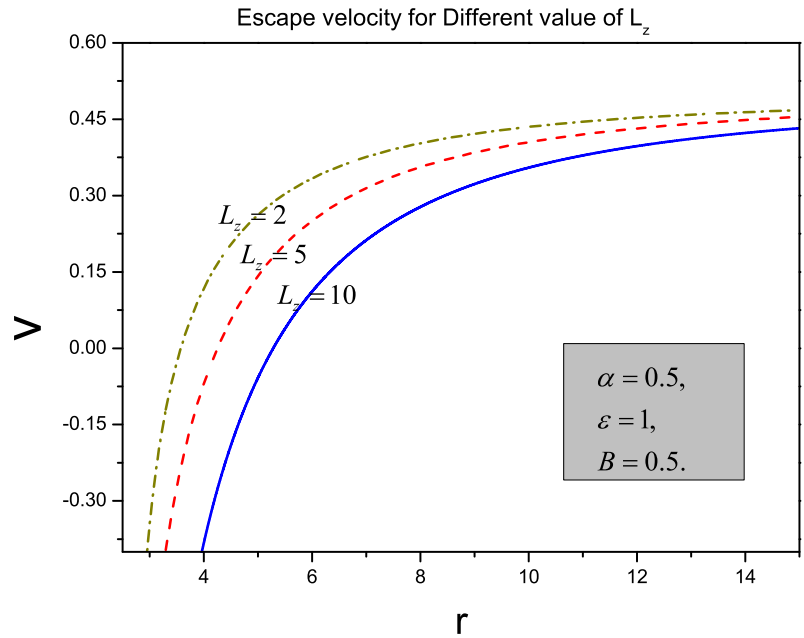}
\caption{Escape velocity ($v$) against $r$ for different
values of angular momentum $\ell$.} \label{fMOG5}
\end{figure}

\begin{figure}[!ht]
\centering
\includegraphics[width=10cm]{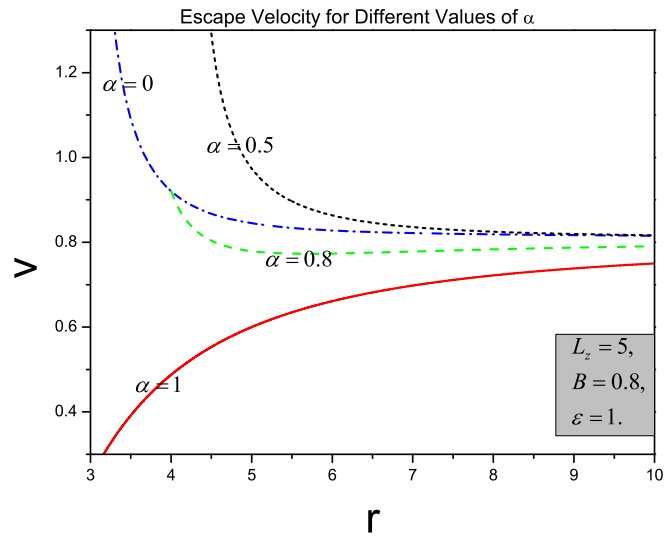}
\caption{Escape velocity ($v$) against $r$ for different
values of parameter $\alpha$.} \label{fMOG9}
\end{figure}

Figure $\ref{fMOG5}$ represents the escape velocity against radius $r$ for different values of angular momentum $L_{z}$.
From figure \ref{fMOG5} we can say that the possibility for a particle to escape having large angular momentum is small.
We have plotted escape velocity for different values of $\alpha$ in figure \ref{fMOG9}
and \ref{fMOG9}, $\alpha=0$, correspond to S-BH (Schwarzschild Black Hole) and $\alpha=1$ correspond to RN-BH.

In figure \ref{fMOG9} we are comparing the escape velocity of a particle
moving around the S-BH, RN-BH and MOG-BH.
Note that the difference between the velocities
is larger  near the black hole (initially) and it becomes almost same  away
from the black hole. Therefore, we can conclude that the effect of the charge
of black hole on the motion of the particle is large while it reduces as particle
moves away from it.
One can see that for large $r$ escape velocity is same for all values of $\alpha$ but
for small $r$ lesser value of $\alpha$ correspond to greater value of escape velocity
and vice versa.
\pagebreak
\section{Summary and Conclusion}

We have investigated  and compare the dynamics of a charged and a neutral particle
in the vicinity of S-BH, RN-BH and MOG-BH.
Geodesics of a neutral and a charged
particle in the vicinity of a MOG-BH are shown in figures
\ref{fMOG18} and \ref{fMOG13}. We see that for a charged particle
there are two boundaries on the geodesics, $r=r_h$
and $r=r_c$ unlike the formal case in which a neutral particle comes
from infinity and goes back to infinity before reaching the horizon,
$r=r_h$, of the BH. We discussed the
effective potential behavior in details regarding the stability of the orbits of the particle.
We further discussed the energy condition for the particle, when it will escape or its motion
remains bound. Expressions for the escape velocity of the particle moving around MOG-BH and for magnetic
field, present in the vicinity of BH due to plasma, is derived in this work.
More importantly a comparison is done for effective
potentials, obtained in the presence and absence of magnetic field.
It is found that presence of magnetic field enhance the stability of
the orbits of the moving particles, due to the presence of it width of
stable region  in contrast to which we
obtained in the case when magnetic field is absent.
We have also done the comparison of the effective potentials
among the RN-BH, S-BH and MOG-BH.
Further we studied the stability by Lyapunov
exponent against magnetic field and a parameter $\alpha$.
We conclude that stability of orbits would increase due to the presence of
vector field considered in MOG.
We deduce that  the particle have to face more repulsion to reach the singularity
due to the presence of vector field as
considered in MOG. But the presence of magnetic field might increase the attractive
force.
It is found that a particle with large value of angular momentum $L_{z}$ would have
greater possibility to reside in the stable orbits in comparison with a particle
with lesser value of it. Therefore, escape velocity  is less correspond to particle with large value
of $L_{z}$.
Effects of
magnetic field and parameter $\alpha$
on escape velocity are also investigated graphically. It is concluded that presence
of magnetic field might provide sufficient energy to particle to escape easily
from the surrounding of BH.
\subsection*{Acknowledgment}
The authors would like to thank both referees and Lim Yen-Kheng for very useful
comments to improve this paper.



\begin{thebibliography}{99}
\bibitem{Dark} C. Deffayet, Phys. Lett. B \textbf{502} (2001) 199;\\
J. S. Alcaniz, Phys. Rev. D {\bf65},  123514 (2002); \\
S. M. Carroll, V. Duvvuri, M. Trodden and M. Turner, Phys. Rev. D \textbf{70}, 043528(2004);\\
S. Nojiri and S. D. Odintsov, Phys. Lett. B 576 (2003) 5; \\
K. Atazadeh and H. R. Sepangi, Int. J. Mod. Phys. D {\bf16}, 687 (2007); \\
S. Nojiri, S. D. Odintsov and M. Sami, Phys. Rev. D {\bf74},  046004 (2006);\\
I. Navarro and K. Van Acoleyen, J. Cosmol. Astropart. Phys. \textbf{0603}, 008 (2006);\\
S. Capozziello, Int. J. Mod. Phys. D {\bf11}, (2002) 483;\\
S. Nojiri and S. D. Odintsov, Phys.Rev. D {\bf68}, 123512 (2003).
\bibitem{rev} S. Capozziello, M. De Laurentis, Phys.  Rept. \textbf{509},
167 (2011);\\ S. Nojiri, S. D. Odintsov, Phys. Rept. \textbf{505},
59 (2011).
\bibitem{55} G. Magnano, M. Ferraris, and M. Francaviglia, Gen. Rel. Grav. \textbf{19}, 465 (1987).
\bibitem{56} J. D. Barrow, and A. C. Ottewill, J. Phys. A: Math. Gen. \textbf{16}, 2757 (1983).
\bibitem{57} N. D. Birrell and  P. C. W. Davies, Quantum Fields in Curved Space  (Cambridge
University Press, 1982).
\bibitem{58} G. Vilkovisky, Class. Quantum Grav. \textbf{9}, 895 (1992).
\bibitem{60} A. A. Starobinsky, Phys. Lett. B \textbf{91}, 99 (1980).
\bibitem{61} D. La, and P. J. Steinhardt, Phys. Rev. Lett. 62, 376 (1989);\\
D. La , P. J. Steinhardt  and E. W. Bertschinger, Phys. Lett. B 231, 231 (1989).
\bibitem{Moffat} J. W. Moffat, Eur. Phys. J. C {\bf75}, 175 (2015).
\bibitem{zak} A. Zakria, M. Jamil, JHEP \textbf{05}, 147 (2015);\\ I. Hussain, M. Jamil, B.
Majeed, Int. J. Theor. Phys. \textbf{54}, 1567 (2015).



\bibitem{Ac} T. Harada and M. Kimura, Class. Quantum Grav. \textbf{31},  243001
(2014).
\bibitem{Null1} M. A. Podurets,  Astr. Zh. {\bf41}, 1090 (1964) (English translation in Sovet Astr.- AJ, {\bf8},868 (1965)).
\bibitem{Null2} W. L. Ames and K. S. Throne, J. Astrophys.  {\bf151}, 659 (1968).
\bibitem{new} C. V. Borm, M. Spaans, Astron. Astrophys. {\bfseries{553}},
L9(2013).
\bibitem{1} J. C. Mckinney,  R. Narayan, Mon. Not. Roy. Astron. Soc. {\bfseries{375}}, 523 (2007).
\bibitem{2} P. B. Dobbie, Z. Kuncic, G. V. Bicknell, and R. Salmeron. Proceedings of IAU Symposium 259:
Cosmic Magnetic Fields: From Planets, To Stars and Galaxies
(Tenerife, 2008).
\bibitem{p9} R. Znajek, Nature {\bfseries{262}}, 270 (1976).
\bibitem{p9a} V. P. Frolov and P. Krtous, Phys. Rev. D \textbf{83}, 024016 (2011).
\bibitem{3} R. D. Blandford, R. L. Znajek, Mon. Not. Roy. Astron. Soc. {\bfseries{179}}, 433 (1977).
\bibitem{4} S. Koide, K. Shibata, T. Kudoh, and D. l. Meier, Science {\bfseries{295}}, 1688(2002).
\bibitem{5} S. Kide, Phys. Rev. D {\bfseries67}, 104010 (2003).
\bibitem{jamil} M. Jamil, A. Qadir, Gen. Rel. Grav. {\bfseries{43}}, 1069 (2011);\\
B. Nayak, M. Jamil, Phys. Lett. B {\bfseries{ 709}}, 118 (2012); \\
M. Jamil, D. Momeni, K. Bamba, R. Myrzakulov, Int. J. Mod. Phys D {\bfseries{21}}, 1250065(2012);\\
M. Jamil, M. Akbar, Gen. Rel. Grav. {\bfseries{43}}, 1061 (2011).
\bibitem{13} A. M. A. Zahrani, V. P. Frolov, A. A. Shoom,  Phys. Rev. D {\bfseries{87}}, 084043 (2013).
\bibitem{saqib} S. Hussain, I. Hussain, M. Jamil, Eur. Phys. J. C \textbf{74}, 3210
(2014).

\bibitem{jamil1} M. Jamil, S. Hussain, B. Majeed, Eur. Phys. J. C \textbf{ 75}, 24
(2015).
\bibitem{Moffat1} J. W. Moffat, JCAP \textbf{0603}, 3004 (2006).

\bibitem{m1} J.W. Moffat, arXiv:1101.1935 [astro-ph.CO]; J. W. Moffat, V. T.
Toth, arXiv:1104.2957 [astro-ph.CO]; ibid, arXiv:1005.2685 [gr-qc];
ibid, AIP Conf. Proc. {\bf1241}, 1066 (2010).

\bibitem{m2} J. W. Moffat, V. T. Toth, Mon. Not. Roy. Astron. Soc. {\bf397}, 1885
(2009).

\bibitem{black} V. Frolov, The Galactic Black Hole (Editors: H. Falcke, F. H. Hehl), IoP (2003).
\bibitem{mag} V. P. Frolov, Phys. Rev. D {\bf 85}, 024020 (2012).
\bibitem{RN} G. Nordstorm, Proc. Kon. Ned. Akad. Wet. {\bf29}, 1238 (1918).
\bibitem{RN1} H. Reissner, Ann. Physik, {\bf50}, 106 (1916).
\bibitem{Kerr} R. P. Kerr, Phys. Rev. Lett. {\bf11}, 237 (1963).
\bibitem{p9b} T.Igata, T. Koike, and H. Ishihara, Phys. Rev. D \textbf{83}, 065027 (2011).
\bibitem{19} L. D. Landau and E. M. Lifshitz, The Classical Theory of Fields (Pergamon Press, Oxford, 1975).
\bibitem{14} S. Chandrasekher, The Mathematical Theory of Black Holes (Oxford University Press, 1983).
\bibitem{punsly} B. Punsly, Black Hole Gravitohydrodynamics (Springer-Verlag, Berlin, 2001).
\bibitem{r1} D. X. Wang , K. Xiao, W. -H. Lei, MNRAS \textbf{335}, 655
(2002).
\bibitem{r2} Wang D.-X., Ma R.-Y., LeiW.-H., Yao G.-Z.,  ApJ \textbf{595}, 109 (2003).
\bibitem{r3} Zhang W.M., Lu Y., Zhang S. N, Chin. J. Astron. Astrophys., \textbf{5}, 347 (2005),.
\bibitem{M} M. Yu. Piotrovich, N. A. Silant, Yu.N. Gnedin, T.M. Natsvlishvili, arXiv:1002.4948v1

\bibitem{10} R. M. Wald, Phys. Rev. D {\bfseries{10}}, 1680 (1974).
\bibitem{6a} A. A. Abdujabbarov, B. J. Ahmedov and N.B. Jurayeva, Phys. Rev. D \textbf{87}, 064042
(2013).
\bibitem{6} A. N. Aliev and N. Ozdemir, Mon. Not. Roy. Astron. Soc. {\bfseries{336}}, 241 (1978).
\bibitem{35} H. Q. Hong, C. J. Hua, W. Y. Jiu Chin. Phys. Lett. \textbf{31}, 060402 (2014).
\bibitem{Fernendo} S. Fernando, arXiv:1202.1502v4

\bibitem{cardoso} V. Cardoso, A. S. Miranda, E. Berti, H. Witeck and V. T. Zanchin, Phy. Rev. {\bf D79}, 064016 (2009).
\bibitem{Lyapunov1} R. Emparan and R. C. Myers, JHEP {\bf0309}, 025 (2003) arXiv: hep-th/0308056.
\bibitem{11} V. P. Frolov and A. A. Shoom, Phys. Rev. D {\bfseries{82}}, 084034 (2010).

\end{thebibliography}
\end{document}